\documentclass[fleqn,usenatbib]{mnras}

\usepackage[T1]{fontenc}

\usepackage{graphicx,psfrag}
\usepackage{mathrsfs}
\usepackage{amsmath,amsfonts,amssymb}
\usepackage{multirow}
\usepackage{comment}
\usepackage{ulem}
\usepackage{hyperref}
\usepackage{enumitem}
\usepackage{tcolorbox}
\usepackage[caption=false]{subfig}

\newcommand{\be}{\begin{equation}}
\newcommand{\ee}{\end{equation}}
\newcommand{\bea}{\begin{eqnarray}}
\newcommand{\eea}{\end{eqnarray}}
\newcommand{\bel}{\begin{align}}
\newcommand{\eel}{\end{align}}

\def\i{{\rm i}}

\def\GMc2{{\rm G M_{\odot} c^{-2}}}

\def\AK{{\tt AthenaK }}

\usepackage{pifont} 

\usepackage{color}
\definecolor{cyan}{rgb}{0,0.9,0.9}
\definecolor{orange}{rgb}{0.9,0.5,0}
\definecolor{magenta}{rgb}{1,0,1}
\definecolor{purple}{rgb}{0.8,0.4,0.8}
\definecolor{gray}{rgb}{0.8242,0.8242,0.8242}
\definecolor{light-gray}{gray}{0.95}
\usepackage{newtxtext,newtxmath}

\title{Turbulence in Magnetised Neutron Stars}
\author[W. Cook et al.]{William Cook$^{1}$ \thanks{Email: william.cook@uni-jena.de},
Raj Kishor Joshi$^{2}$,
Sebastiano Bernuzzi$^{1}$,
Brynmor Haskell$^{2,3,4}$
and Jacob Fields $^{5,6}$
\\
$^{1}$ Theoretisch-Physikalisches Institut, Friedrich-Schiller-Universit{\"a}t Jena, 07743, Jena, Germany\\
$^{2}$Nicolaus Copernicus Astronomical Center of the Polish Academy of Sciences, Bartycka 18, 00-716, Warsaw, Poland\\
$^{3}$INFN, Sezione di Milano, Via Celoria 16, 20133, Milano, Italy\\
$^{4}$Dipartimento di Fisica, Universit\`{a} di Milano, Via Celoria 16, 20133, Italy\\
$^{5}$Institute for Gravitation and the Cosmos, The Pennsylvania State University,
  University Park, PA 16802, USA\\
$^{6}$Department of Physics, The Pennsylvania State University, University Park,
  PA 16802, USA\\
}
\begin{document}
\label{firstpage}
\pagerange{\pageref{firstpage}--\pageref{lastpage}}
\maketitle

\date{\today}

\begin{abstract}
The magnetic field configuration in the interior of neutron stars and its stability are  open problems and may be impacted by the influence of a turbulent cascade within the star. Assessing the impact of turbulent flow with numerical simulations requires incredibly high resolution as well as long lived simulations covering multiple Alfven times. We present a series of simulations of magnetised neutron stars with resolution up to 29m and lasting at their longest 1.2s to assess this issue, the longest lasting and highest resolution such simulations to date.  At the highest resolution we find evidence for a turbulent cascade absent in an unmagnetised star which cannot be captured with lower resolution simulations, consistent with Kolmogorov power law scaling. The presence of turbulence triggers an inverse cascade of helicity, while at late times the net helicity appears to vanish, suggesting that a twisted-torus is not formed in the magnetic field. We find that the presence of the magnetic field excites a characteristic quadrupolar oscillation of the density profile at $145$Hz, consistent with Alfvenic modes proposed as the source of quasi-periodic oscillations observed in magnetars.
\end{abstract}


\section{Introduction}
\label{sec:intro}

Neutron Stars (NSs) are strongly magnetised compact objects, with inferred dipolar
magnetic fields at the surface that range from $B\approx 10^8$ G in
older systems such as the millisecond pulsars, to $B\approx 10^{16}$ G
in magnetars, for which the magnetic energy is a significant
fraction of the total energy of the star and is thought to power the
observed outbursts \citep{Thompson:1995gw}.

To date constraints on the field structure close to the star have been obtained mainly from hotspot modelling of
four millisecond X-ray pulsars observed by the Neutron Star Interior
Composition Explorer (NICER) \citep{Gendreau:2016}, which
reveals that the magnetic field is generally not a simple dipole, but
has to be modelled with a more complex multipolar structure or an
offset dipole \citep{Bilous:2019,Riley:2021pdl,Choudhury:2024,Salmi:2024}.

Observations of so-called `low field' magnetars, systems which exhibit
magnetar-like outbursts but have low magnetic fields inferred from
spindown \citep{Rea:2010}, also suggest that strong toroidal or
small scale components of the field are required close to the
surface \citep{Igoshev:2021,Tiengo:2013,RodriguezCastillo:2016}.

The possibility that the interior field may be significantly stronger
than the inferred exterior dipole is particularly interesting for
Gravitational Wave (GW) searches, as it would imply that some NSs may
be significantly deformed by their magnetic fields and emit a
detectable continuous signal \citep{Bonazzola:1995rb}.

It is well known that the interior field of a NS cannot be purely
poloidal or purely toroidal, as such configurations are unstable to the varicose and kink instabilities
\citep{Tayler:1957a, Tayler:1973a, Markey:1973a, Markey:1974a,
  Wright:1973a}, but must rather be a mixture of the two components in
what is generally assumed to be a twisted torus
\citep{Braithwaite:2005xi}. The toroidal component, in particular, may
be significantly stronger than the poloidal, leading to a low inferred
external field but strong effects on the bursting activity and on GW
emission \citep{Braithwaite:2009}.

While equilibrium models of twisted-tori with strong toroidal
components can be constructed \citep{Ciolfi:2013dta}, it is however
unclear if such configurations are stable, and it has in fact, been
suggested that no stable configurations exist in barotropic NS models
\citep{Lander:2012a}, and that stratification effects
\citep{Becerra:2022} or elasticity \citep{Fujisawa:2023,Bera:2020} 
 must be included to
stabilise the field.

In order to investigate this issue a number of time evolutions of
magnetised NSs have been carried out, both in Newtonian gravity
\citep{Braithwaite:2005md,Sur:2020hwn} and in General Relativistic MagnetoHydrodynamics (GRMHD) \citep{Kiuchi:2008ss,Ciolfi:2011xa,Lasky:2011un,Lasky:2012ju,Ciolfi:2012en,Sur:2021awe,Tsokaros:2021pkh,Cook:2023bag,Cheong:2024stz}. Such simulations can describe the first few hours of a neutron star's life after cooling, before the matter becomes superfluid, and the ideal MHD approximation is valid, after which the field becomes frozen in. The problem of determining
whether a stable equilibrium develops is, however, complex and
computationally demanding. 
In static stars, with no dynamo effects amplifying the field strength, 
numerical dissipation causes magnetic energy to be lost in simulations, leading to an increase in the Alfven
timescale that governs the dynamics of the field, that requires long
lived simulations, on the other, as instabilities of the field
develop, non-ideal effects play a strong role and turbulence
develops. To accurately capture its effect on the development of large
scale structures in the field, however, high resolution runs are
necessary.

This is essentially the conclusion of \cite{Sur:2021awe}, who
performed long lived simulations of an initially poloidal field,
finding that a toroidal field, contributing to roughly $10\%$ of the
total field, was generated, but also observed the development of turbulence. 
 At the resolutions considered however, comparison with an unmagnetised star showed that the spectra became identical, suggesting that the turbulence was triggered not by the magnetic field, but by the perturbation arising from the numerical simulation.

\cite{Sur:2020hwn} found that non-ideal
MHD effects lead to a violation of helicity conservation, 
which measures the `twist' of the magnetic field, and that
turbulence leads to an inverse cascade of helicity, transferring
energy from small resistive scales to the larger scale toroidal
components. This is a crucial issue not only for the development and
stability of the magnetic field in older pulsars, but may also be
linked to the development of strong magnetic fields at the star's birth.
Simulations of helical turbulence in a box suggest that the inverse
cascade can be strong enough to develop a large scale field from
initially random fields on small scales, which would lead to the
possibility that NS magnetic fields are not, fossil, but entirely
generated by turbulence \citep{Sarin:2023}.

The nature of the spectrum associated to MHD turbulence is still unclear, with key predictions of scaling behaviours made by \cite{Kolmogorov:1941,Iroshnikov:1964,Kraichnan:1965,Goldreich:1995}. Details of theoretical and computational progress on MHD turbulence can be found in reviews of \cite{Beresnyak:2019,Schekochihin:2020aqu}.
The modeling of turbulence in NSs also plays an important role in the study of binary NS mergers, see \cite{Radice:2024gic} for a recent review.

In summary, previous studies have proven inconclusive as to whether magnetic fields can drive a turbulent cascade within a NS, which may itself cause a large scale rearrangement of the magnetic field, due to limited resolution.
In this paper we report a new set of the highest resolution, longest lasting, simulations to date, which allow us to resolve the
magnetic effects on the initial development of turbulence, and study
the dynamical stability and topology of the field and its helicity.

\section{Method}
\label{sec:methods}

We perform a sequence of simulations of a static NS with and without magnetic fields using the code 
\AK \citep{Stone:2024,Zhu:2024utz,Fields:2024pob}. We perform our evolutions in the Cowling approximation, holding the spacetime metric fixed as that of the TOV star described below. 
The computational domain consists
of a block based statically refined grid with the NS wholly covered by the finest refinement level.
Runs are performed at four resolutions in a convergent series, and are referred to by
the grid spacing $\delta x$ on the finest refinement level covering the star, in units
of metres. The resolutions considered are: $(230,115,57,29)$m. Runs denoted by $B0$
are reference runs without a magnetic field. At the highest resolution ($\delta x = 29$m) 
we simulate for 100ms of evolution, while at the lowest resolution ($\delta x = 230$m) we evolve for 1.2s.

To date the longest lasting simulations of isolated neutron stars have been performed for 880ms in \cite{Sur:2021awe}, while the highest resolution simulations were performed by \cite{Tsokaros:2021pkh} with a resolution of  72m, evolved for $\sim3$ ms; these runs therefore constitute the longest and highest resolution of such simulations. We comment on the comparison between these runs in terms of Alfven time in Appendix \ref{app:alf}. 
We place the outer boundaries of the computational domain at 59km from the origin, see Appendix~\ref{sec:BC} for a study of outer boundary effects.

For initial data we consider a TOV star with central density $7.91\times10^{14}$g(cm)${}^{-3}$ and radius 12.0km, with a Gamma law EOS such that $p = \rho \epsilon (\Gamma -1)$, with $\Gamma = 2$. This is the commonly considered model A0 of \cite{Dimmelmeier:2005zk}, corresponding to a non-rotating NS without stratification. We construct this model by solving the TOV equations with a shooting method, targeting the given central density. Outside the star, the computational domain is filled with a low density atmosphere fluid, fixed at the density $\rho_{\mathrm{atm}} = 62$ g(cm)${}^{-3}$. 
In the initial data we set the fluid pressure $p$ using the barotropic EOS $p = K \rho^\Gamma$, where $K=100, \Gamma=2$. This EOS is also used to set the pressure in the atmosphere.

We initialise the magnetic field with a purely toroidal vector potential, 
\begin{eqnarray*}
	A^\phi = A_b\max(p,p-0.04p_{\mathrm{max}}) 
\end{eqnarray*}
setting parameter $A_b$ to achieve an average magnetic field strength of $1.23\times 10^{16}$G  in the star. This is consistent with magnetar field strength, and those previously studied by the authors in \cite{Sur:2021awe,Cook:2023bag}.
$p_\mathrm{max}$ is the maximum value of the pressure in the star interior. The magnetic field is then determined by $\mathbf{B} = \nabla \times \mathbf{A}$. In our simulations we do not seed an initial perturbation, rather the instability is triggered by truncation error from the numerical solution of the equations on a Cartesian grid. In our evolutions we employ the PPMX reconstruction of \cite{Colella:2008}, and use the HLLE  approximate Riemann solver for flux calculation \citep{Harten:1983,Einfeldt:1988} augmented with a first order flux correction as detailed in \cite{Fields:2024pob}. The magnetic field evolution is performed using the Constrained-Transport algorithm of \cite{Evans:1988a}, as implemented in \AK following \cite{Gardiner:2005,Gardiner:2007nc}. Time evolution is performed using the strong stability preserving third order Runge-Kutta scheme of \cite{Gottlieb:2009a} with Courant-Lewy-Friedrich factor 0.5.

\section{Results}
\label{sec:res}

\subsection{Magnetic Field Dynamics}
\label{sec:dynamics}

\begin{figure*}
  \centering 
    \subfloat[$t=12.5$ms]{
    \includegraphics[width=0.3\textwidth]{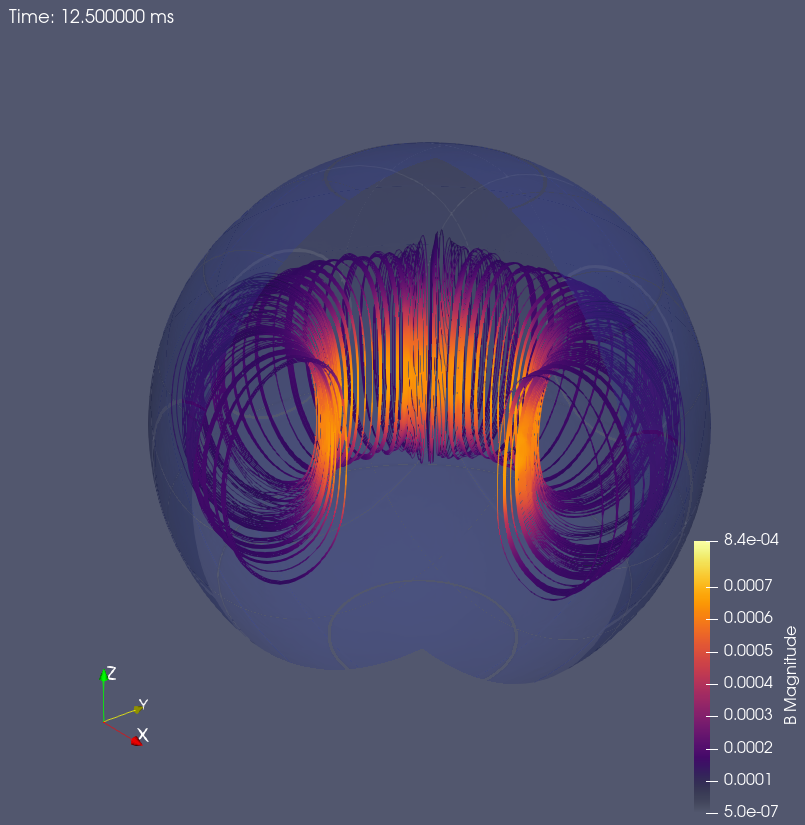}
  \label{fig:stream_a}
}
	\subfloat[$t=25$ms]{
    \includegraphics[width=0.3\textwidth]{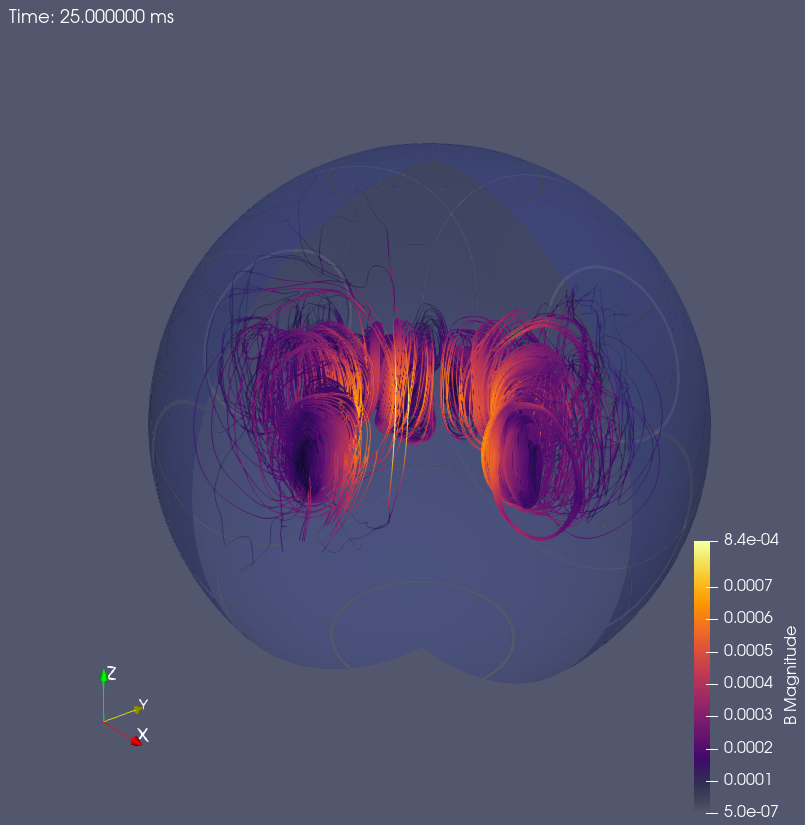}
  \label{fig:stream_b}
}
 \subfloat[$t=75$ms]{
    \includegraphics[width=0.3\textwidth]{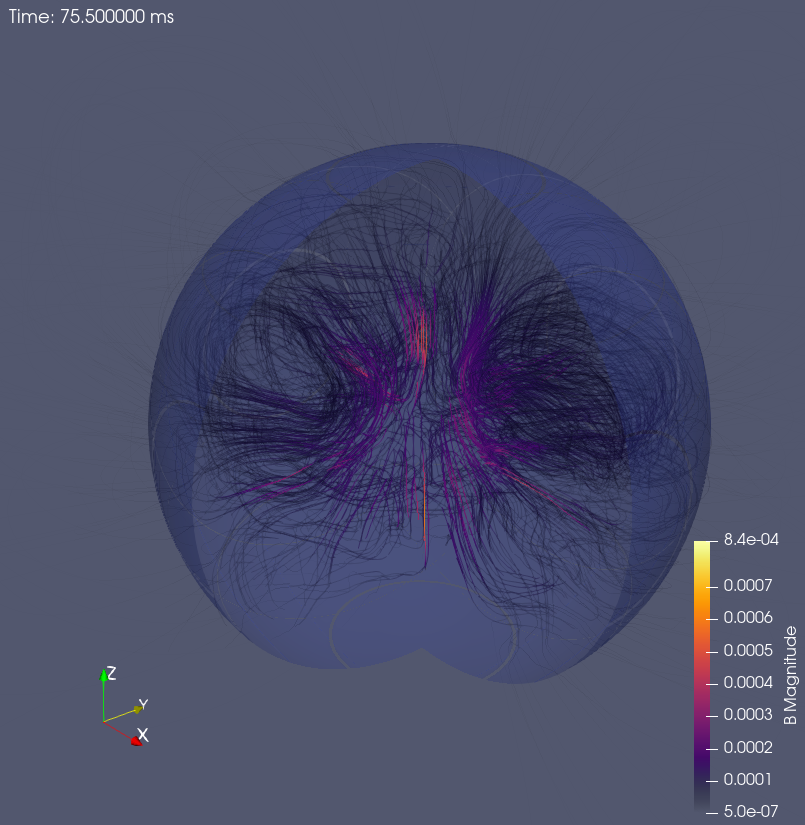}
  \label{fig:stream_c}
}

	  \caption{Magnetic fieldlines for the $57$m magnetised
            simulation. Snapshots correspond to:
            (a) $t=12.5$ms, saturation of the varicose instability, and onset of kink instability.
            (b) $t=25$ms, saturation of the kink instability.
            (c) $t=75$ms, late time configuration.
            The streamlines are seeded on a circle of radius $4.4$km
            with their colour and opacity showing the strength of the
            field. 
            The translucent isocontour corresponds to $1\%$ of the
            initial maximum density. 
          }
          	  \label{fig:stream}
	\end{figure*}

We first discuss the qualitative evolution of the magnetic field configuration. As expected, the initial poloidal field is subject to instabilities, which we follow the evolution of in Fig. \ref{fig:stream}.

Fig. \ref{fig:stream_a} illustrates the development of the varicose instability, as the cross sectional area of the field lines loses axisymmetry, picking up a clear $m=4$ azimuthal dependence, as seeded by the grid structure. At this time small scale eddy features begin developing in the magnetic field, with associated toroidal fields of magnitude comparable to the initial average magnetic field strength. Fig. \ref{fig:stream_b} shows the development of the kink instability, as field lines are displaced orthogonal to the radial gravitational field. By late times, Fig. \ref{fig:stream_c} demonstrates that the initial global poloidal field structure has largely dissipated, with the magnetic field amplitude greatly reduced and the field configuration itself much more chaotic, with little visible overall structure. At this point the eddies are still visible, with similar local magnitudes to early times.

\subsection{Energetics}
 \label{sec:energy}
We now discuss the evolution of the energy budget of the star, defined in App.~\ref{app:energies}, 
demonstrating the evolution of the star towards an energetically steady state.

\begin{figure}
	\centering 
	\includegraphics[width=\columnwidth]{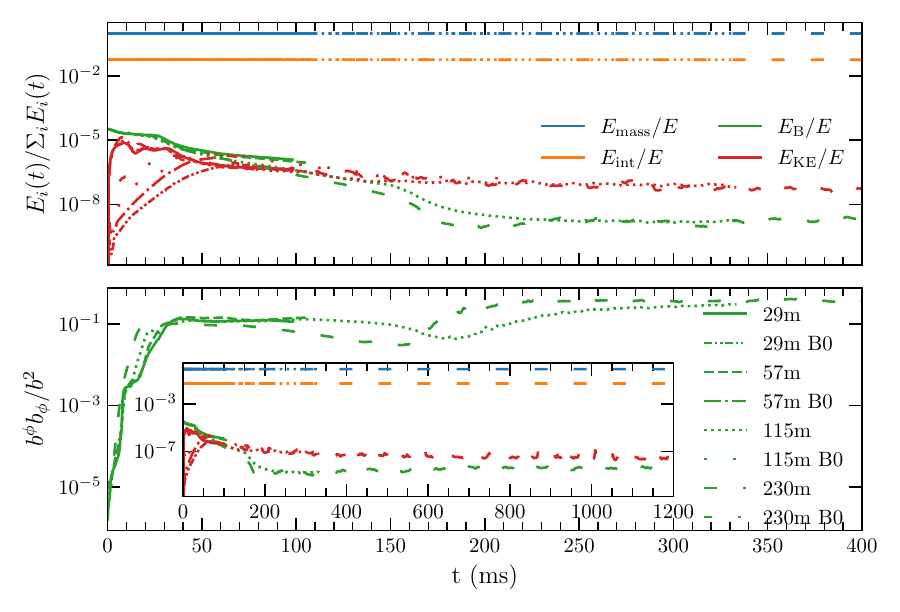}
	\caption{Evolution of energies in magnetised and unmagnetised
          configurations. (Upper) The proportion of the total energy
          given by each sector. (Lower) The evolution of the toroidal
          component of the magnetic field normalised by the total
	  magnetic field strength. (Inset lower) A zoom out of the upper panel showing 1.2s of evolution.}
	\label{fig:energy}
\end{figure}

In Fig. \ref{fig:energy} we demonstrate the evolution of energy in the NS over the first 400ms
of evolution for our simulations with and without a magnetic field for all four resolutions. 
In the upper panel of Fig. \ref{fig:energy} we see the contribution to the total energy budget 
from various sectors. The energy of the spacetime is dominated by the rest mass density and the 
contribution from the internal energy of the fluid throughout the evolution in all configurations. 
Let us now focus on the quantitative behaviour of the highest resolution run, which behaves similarly to the other resolutions over the first 100ms.

  The initial magnetic field energy is $0.003\%$ of the total energy; the
  initial kinetic energy is zero as the NS is a static solution to  
  the Einstein equations. The kinetic energy rapidly grows driven by
  the rearrangement of the magnetic field, which undergoes the onset of the instabilities described 
  in Sec.~\ref{sec:dynamics}.
  In the ideal MHD approximation the fluid streamlines are pinned to 
  the magnetic field lines, and so, as the field changes, the fluid must be dragged along with the 
  field lines, leading to a growth in kinetic energy, and a corresponding decrease in magnetic field 
  energy. 
  The kinetic energy peak at ${\sim} 9$ms corresponds to the
  saturation of the varicose instability; at this point the kinetic
  energy is ${\sim} 0.001\%$ of the total energy.
  Afterwards, the kink instability takes over and saturates at about
  ${\sim}30$ms. As the kink saturates, energy is exchanged between the
  magnetic and kinetic sectors; the magnetic energy drops to
  ${\sim}25\%$ of its initial value by the ${\sim}30$ms mark.
  In the nonlinear phase, both magnetic and kinetic
  energies decrease steadily, with 
  the magnetic energy dropping to ${\sim}3\%$ of
  its initial value by ${\sim}100$ms. We ascribe this to energy loss as small
  scale structures are lost from the grid due to finite resolution
  effects. We comment further on the formation of these structures in
  Sec.~\ref{sec:spectra}. 

Our simulations show a good conservation of energy, with the total energy conserved to one part in 
$10^5$, internal energy conserved at one part in $10^4$ and rest mass conserved at one part in 
$10^7$. We do not expect a perfect conservation of internal energy, as the dynamics of the fluid will lead to an exchange in energy between the kinetic, magnetic and internal energy sectors, however, we note that this conservation of energy and in particular internal energy is a considerable 
improvement on previous results in \cite{Sur:2021awe}. This is due to a combination 
of an improved atmosphere treatment and conserved-to-primitive approach, detailed in the appendix of 
\cite{Cook:2023bag}; extended outer boundaries (see App. \ref{sec:BC}); and the first order flux correction implemented in \AK, detailed in 
\cite{Fields:2024pob}.   A slight rest-mass loss occurs at ${\sim}12$ms, due to low density matter leaking 
off the surface of the star exiting the outer boundary of the computational domain.

In the lower panel of Fig. \ref{fig:energy}  we show the growth of the toroidal component of the 
magnetic field, as expected, to stabilise the configuration. 
Up until ${\sim}9$ms we see 
a rapid growth of the toroidal component, growing to $\sim 0.2\%$ of the total magnetic field 
energy, corresponding to the initial varicose instability. Subsequently the growth rate becomes more 
gradual as the kink instability begins to dominate. After $\sim 30$ms the toroidal energy has reached 
$\sim 10\%$ of the magnetic field energy, peaking at $15\%$ after ${\sim}41$ms. 
For the remainder of the simulation this component oscillates between $12$ and $15\%$.

In the upper panel of Fig. \ref{fig:energy}  we include also fiducial runs with no magnetic field at
the same resolutions.
  The key difference between the MHD and hydrodynamics simulations is
  the behaviour of the kinetic energy. Without a magnetic field
  driving the fluid motion,
  the growth of kinetic energy is suppressed.
Instead, fluid motion is triggered by truncation error in the simulation, most 
prominent at the sharp surface of the NS. This kinetic energy slowly grows over time, 
peaking at ${\sim}60$ms and slowly decreasing afterwards due to finite resolution effects.

In the inset of the lower panel of Fig. \ref{fig:energy} we show the energy evolution for our full 
series of runs at varying resolutions and at far longer timescales. 
Firstly we note that the magnetic and kinetic energies
are consistent over different resolutions, as is the growth of the toroidal component of the field. 
The initial growth of the toroidal component is delayed at higher resolutions, 
since the truncation errors which trigger the
   magnetic-field instabilities are smaller.
Further, the growth of the kinetic energy in the unmagnetised
  cases is slower at higher resolutions due to smaller and convergent truncation errors.
These results are consistent with behaviour seen in
previous simulations at lower resolutions, for shorter timespans, in
\cite{Cook:2023bag}, albeit that those
are runs with a fully dynamical spacetime.

We now comment on the late time behaviour that we see
with lower resolution simulations. After 1200 ms of
evolution, conservation of energy and of rest mass are maintained to the level
of 0.03\% and $8\times 10^{-5}\%$ respectively.
After $\sim 200-300$ms the
energetics of the star appear to settle down into a quasi-equilibrium
configuration, with the magnetic field energy ending at $4\times
10^{-7}$\% of the total energy and the kinetic energy at $2\times
10^{-6}$\%. We see the toroidal field grow slightly between 150 and
300ms, and after this time oscillate between 34\% and 42\% of the
total magnetic field energy, ending at a value larger than that found
at the end of the 100ms evolution of our highest resolution run. 
These values should be confirmed with yet longer higher resolution
simulations. Due to the dissipation of the magnetic field, the Alfven
time increases considerably through the course of the simulation so that a slow evolution
of the field should be expected at late times (see App. \ref{app:alf}).

\subsection{Mode analysis}

The profile of the magnetic field that forms
after a long term evolution of the linear instabilities and non linear evolution of the field 
discussed in Sec. \ref{sec:dynamics} is currently poorly constrained.
This profile will however cause a deformation of the
initially spherical star through the exertion of magnetic pressure. 
Such deformations will excite oscillations of the 
stars density profile with the characteristic frequencies of the star \citep{Stergioulas:2003ep,Baiotti:2008nf,Leung:2022mvm} and
may lead to ellipticities which 
can source long-lived continuous gravitational wave signals \citep{Bonazzola:1995rb}. To characterise this 
profile we measure the radially integrated angular distribution of three quantities, the density, 
$\rho$, the toroidal field $B^\phi$ and the magnetic field strength $B^2$. We analyse a full
decomposition into spherical harmonics, as well as a decomposition into purely azimuthal modes by expressing a quantity of interest, $f$, either as $
f = \sum_{\ell,m} a_{\ell,m}(r) Y_{\ell,m}(\theta, \phi),
$
or as 
$
f = \sum_{m} c_{m}(r,\theta) e^{ i m \phi}.
$

\begin{figure}
	\centering 
	\includegraphics[width=\columnwidth]{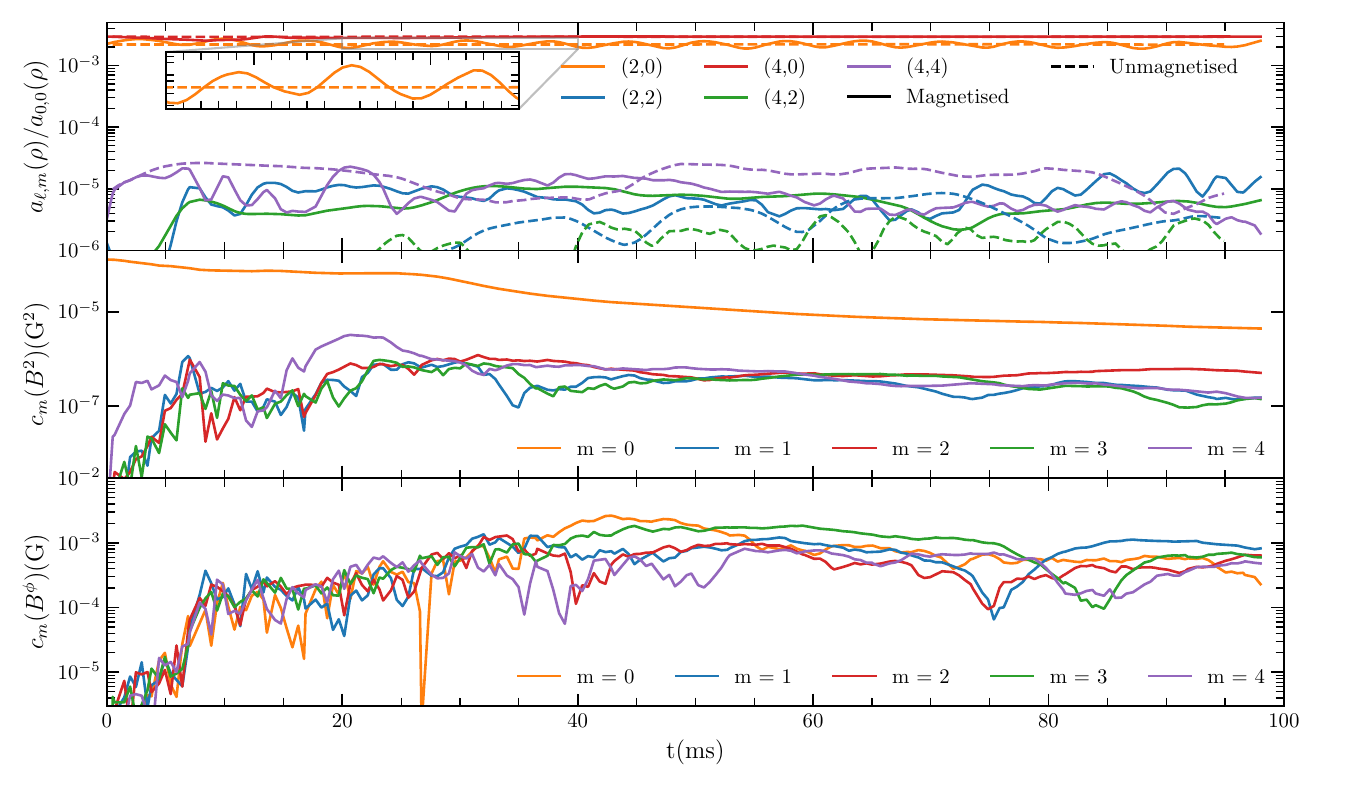}
	\caption{Mode decomposition for 29m resolution. Upper panel: Evolution of spherical harmonic mode contributions to density
          profile. Solid lines
          represent the run with magnetic fields, dashed lines
          represent the unmagnetised configuration.  Modes are normalised by the coefficient $a_{00}$,
	  which is the overall dominant mode. Inset: Zoom in on the (2,0) mode. Middle panel: Evolution of azimuthal mode contributions to overall magnetic field strength $B^2$. Lower panel: Evolution of azimuthal mode contributions to toroidal magnetic field $B^\phi$.} 
	\label{fig:modes}
\end{figure}

We demonstrate the mode decomposition in Fig. \ref{fig:modes}.
In the upper panel we investigate the decomposition of the density profile 
into spherical harmonics for the unmagnetised and magnetised
configurations.
From hereon we refer to a given mode by the pair $(\ell,m)$. We plot just the 
dominant modes $(2,0),(2,2),(4,0),(4,2),(4,4)$. Of the non axisymmetric modes 
the $(4,4)$ mode mostly dominates in magnitude, as expected due to the nature of the Cartesian grid 
seeding of the instability (c.f. Fig \ref{fig:stream_a}).  While this mode begins the largest in 
both the magnetised and unmagnetised run, in the magnetised case it considerably decreases compared 
to the unmagnetised run during the first ${\sim}25$ms while the linear instabilities grow. This 
corresponds to a considerable increase in all the other modes with
$m=1,2,3$ over this initial phase, as the magnetic field rearranges,
as can be seen here in the $m=2$ sector. After this, the  
modes remain at similar levels with and without a magnetic
field. 
By the end of the simulation it is clear that, while in the unmagnetised case it is the $m=4$ mode that dominates, in the magnetised case the amplitude is more evenly divided into the $m=2$ contribution also.

We now focus on the axisymmetric sector. In the 
unmagnetised case $a_{2,0}$ takes an approximately constant value throughout 
the simulation. In contrast, the magnetised case oscillates around this value with a clear 
characteristic frequency of ${\sim}145$Hz. Perturbative calculations of Alfvenic oscillations in 
magnetised NSs suggest that oscillation frequencies in this range can be expected to be excited by 
magnetic fields, though we note that the precise value of these oscillations depend on the magnetic 
field model considered, and on the modelling of the NS crust \citep{Sotani:2007pp,Cerda-Duran:2009hdu,Sotani:2009pg,Lander:2011}. These oscillations have been proposed 
as an explanation for the quasi-periodic oscillations (QPOs) observed in pulsar observations, with 
the 150Hz signal in the X-ray tail of SGR 1806-20 \citep{Israel:2005av} proposed as an overtone of such a mode \citep{Sotani:2007pp}. Such modes 
are expected to scale linearly with the magnetic field strength, and, by performing low resolution 
simulations with an increased magnetic field amplitude, we see tentative evidence of higher frequency 
modes excited. There exists extensive discussion as to whether such modes should form a discrete or continuum spectrum, varying as a function of the location in the star~\citep{Levin:2006ck,Levin:2006qd,Glampedakis:2006apa}. We re-calculate $a_{2,0}$ at a variety of radii within the star without the radial integration applied, and find a consistent value of $145$Hz at all radii within the star, suggesting that this is a discrete mode frequency. It is known that in the Cowling approximation that the observed frequencies of such modes in NSs can be altered by a factor ${\sim}2$ \citep{Font:2001ew}. A precise measurement of this frequency will require further simulations with a fully dynamical metric.

The distribution of magnetic field energy within the star is
  demonstrated in the middle panel of Fig. \ref{fig:modes}.
The initial poloidal 
configuration is axisymmetric, and so we see the $m=0$ mode
dominate at early times.
As the toroidal component grows and instabilities develop, this
symmetry is quickly broken, and $c_0$
decays throughout the simulation. As with other field
components, as the initial instabilities grow, higher
modes grow over the first $\sim 20$ ms with the $m=4$ contribution dominating. The higher mode
contributions ($m=2,3,4$) have peaked by ${\sim}40$ ms and subsequently
decay, with the final configuration dominated by the $m=0$ mode.
As expected, due to the grid imprint, the $m = 2$ and $m = 4$ modes are the next most dominant,
though we note that this feature is resolution dependent, and that at lower resolutions the $m = 1$ mode becomes the second strongest contribution.

In the lower panel of Fig. \ref{fig:modes} we demonstrate the azimuthal dependence of the toroidal field $B^\phi$. Since in a non rotating star there is no preferred direction to construct a coherent toroidal field given by angular momentum, 
we do not expect the $m=0$ mode to dominate.
Initially the field is zero, and during the linear instability dominated phase up until $\sim 40$ms we see all modes growing at approximately the same rate. Subsequent to this phase the $m=0$ mode begins to consistently 
dominate until 50 ms, decaying after ${\sim}45$ms. At late times we see an interchange between the $m = 3$ and $m = 1$ modes, though at lower resolutions we see a dominant $m = 0$ mode.
When analysing the full spherical harmonic decomposition, we see a similar growth in all modes at early time in terms of both amplitude and growth rate, as the toroidal field develops. By $\sim$ 30 ms this growth has saturated, with contributions at larger $\ell$ and larger $m$ slightly dominating. These effects begin to decay past the $50$ms mark.

\subsection{Turbulence}
\label{sec:spectra}

\begin{figure*}
		\centering 
		\subfloat[$t=2.995$ms]{
		\includegraphics[width=0.49\textwidth]{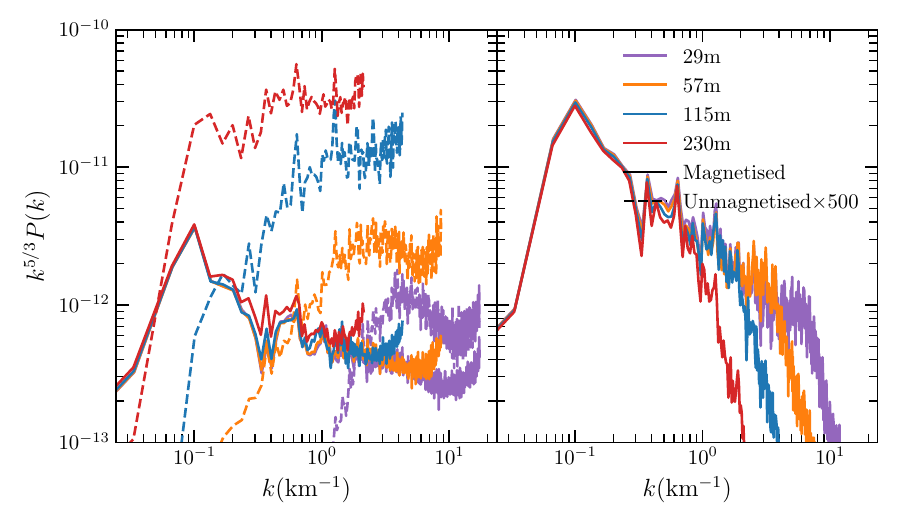}
		\label{fig:powerspec_a}
			}
\subfloat[$t=34.478$ms]{
	\includegraphics[width=0.49\textwidth]{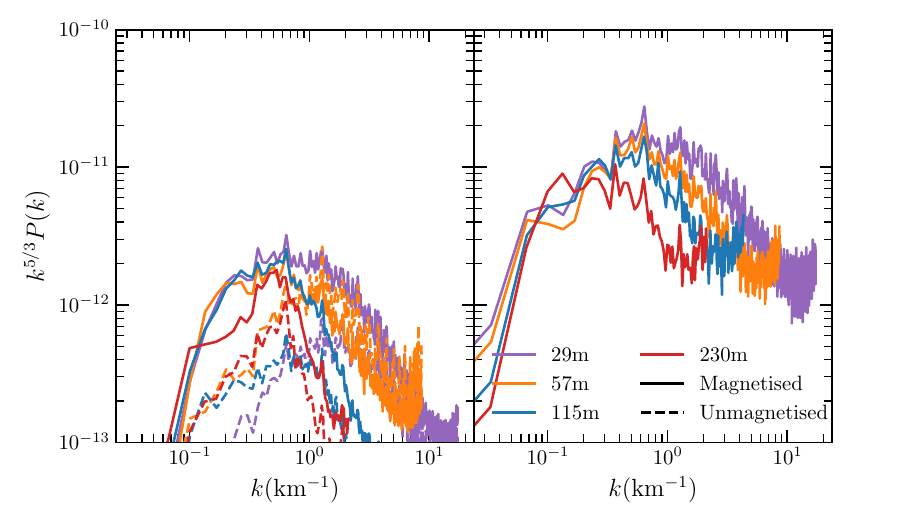}
	\label{fig:powerspec_d}
}\\
\subfloat[$t=60.091$ms]{
	\includegraphics[width=0.49\textwidth]{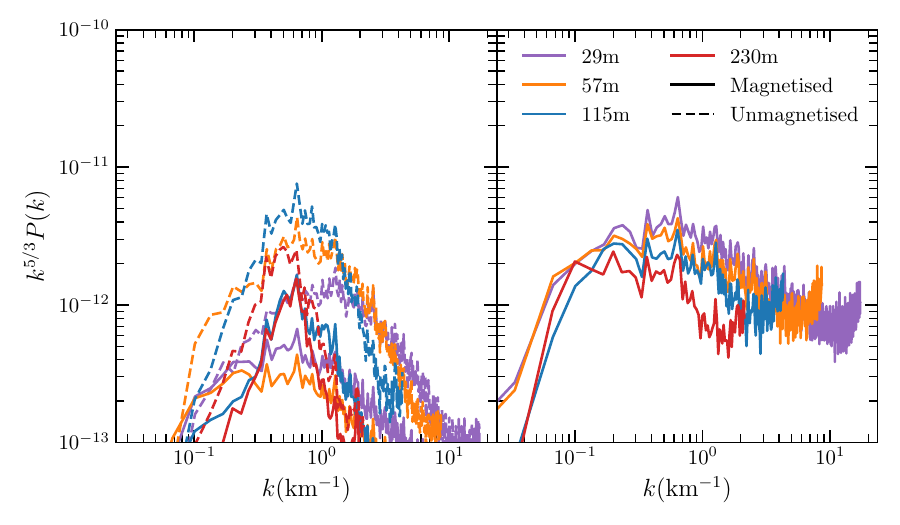}
	\label{fig:powerspec_f}
}
\subfloat[$t=78.315$ms]{
	\includegraphics[width=0.49\textwidth]{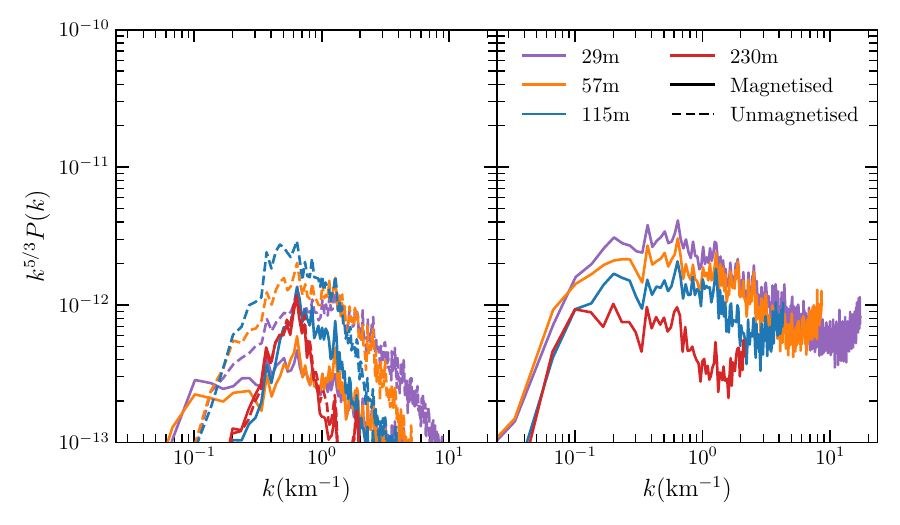}
	\label{fig:powerspec_g}
}
	\caption{Spectra of the kinetic energy (left) and magnetic energy (right) rescaled by $k^{5/3}$ for various times. Solid lines represent magnetised configurations, while dashed lines represent unmagnetised configurations. Colours denote different resolutions. In panel \ref{fig:powerspec_a} only, the unmagnetised kinetic spectra is magnified by a factor of 500 for visibility.}
		\label{fig:powerspec}
	\end{figure*}

At high Reynolds numbers inside a NS the onset of turbulent flow as the response to a perturbation, such as the evolution of the magnetic field instability, may be expected
\citep{Radice:2024gic} . 
The previous results of \cite{Sur:2020hwn}
have suggested that, in Newtonian MHD simulations similar to those
presented in this paper albeit at lower resolutions, a turbulent
cascade can be seen in the kinetic and magnetic power spectra, with
some suggestion that the scaling behaviour follows either the
predicted $k^{-5/3}$ scaling of  \cite{Kolmogorov:1941}
(K41) or the $k^{-3/2}$ scaling of 
\cite{Iroshnikov:1964,Kraichnan:1965} (IK). Subsequent simulations at
higher resolution in GRMHD with the Cowling approximation
\citep{Sur:2021awe} made a comparison with a fiducial unmagnetised
run, and found that the spectra became identical between the two
cases, suggesting that the turbulent behaviour was triggered not by
the presence of the magnetic field, but simply by the perturbation
applied to the fluid arising from the numerical simulation. Here we address the origin of this turbulent behaviour.

We demonstrate the time evolution of the kinetic and magnetic power spectra (defined in App. \ref{app:specdef}) in Figure \ref{fig:powerspec}. All power spectra are rescaled by a factor of $k^{5/3}$, consistent with the Kolmogorov scaling factor. Therefore a flat feature in Fig. \ref{fig:powerspec} corresponds to the turbulent cascade.

At early times (Fig. \ref{fig:powerspec_a}), while the linear instabilities begin to grow,  for the magnetised configurations both kinetic and magnetic spectra are dominated by long wavelength features due to the dynamics of the large scale field. In the unmagnetised case  the kinetic spectrum converges to zero, dominated by small wavelength features at the star surface. 

As the linear instability develops, as seen in Fig. \ref{fig:energy}
the kinetic energy grows, before peaking at $\sim 10$ms and eventually
beginning to decay. By $t{\sim}35$ms 
(Fig. \ref{fig:powerspec_d}) the kink instability and the toroidal field growth have saturated, and for the first time in the evolution we observe a flat feature corresponding to a $k^{-5/3}$ power law scaling forming in the spectra. This feature is clearest in the 29m simulation in the kinetic spectrum, with the feature less obvious at lower resolutions. In the unmagnetised case, the kinetic power spectrum is still much weaker, and there is little evidence for a flat feature in the spectrum. In the magnetic field spectrum the evidence for turbulence is less strong at this point in time, with a much narrower flat range visible within the spectrum. 

As the fully nonlinear phase develops the $k^{-5/3}$ feature in the 29 and 57m kinetic spectra is visible, becoming noticeably clearer after $\sim 45$ms, with a similar feature developing in the magnetic spectrum at a similar time. By 60 ms (Fig. \ref{fig:powerspec_f}) this feature is very strong in the 29 and 57m kinetic spectra. At this point it is clear that there is a turbulent spectrum in the magnetised star which is not present in the corresponding unmagnetised configuration, within which there is little evidence for a scaling feature in the cascade. This feature must be driven by the magnetic field dynamics, rather than purely the response of the fluid to numerical perturbation. It is also clear at this point that the 29 and 57m simulations are capturing behaviour absent from the 115m and 230m simulations, for which the resolution is insufficient to capture the width of this cascade in terms of wavenumber. In the magnetic spectrum a flat feature is also forming, captured best at the highest resolutions also. 

In Fig. \ref{fig:powerspec_g} we observe the power spectra close to
the end of the highest resolution runs. Here we see that, even on very
long timescales, the turbulent spectrum is still relatively well
captured at the 29 and 57m resolution in both the kinetic and magnetic
spectra. By this time, we see that at the lowest resolution, 230m, the
kinetic energy spectrum has become identical between the magnetised
and unmagnetised runs, which show no signs of a turbulent
cascade. At 
low resolution or for sufficiently long evolution times, the impact of the presence of a magnetic field on the velocity field will eventually be washed out by the accumulation of numerical error driving the perturbation of the NS. These are the features previously seen in e.g. \cite{Sur:2021awe}. As we continue these simulations to later times we see that the 230m kinetic spectra with and without a magnetic field remain close, while after ${\sim}$ 150ms of evolution the 115m kinetic spectra also begin to match. It is reasonable to assume that, for a sufficient evolution time, we would see the same behaviour in the 29 and 57m configurations. We note that this behaviour does not occur as soon as the unmagnetised kinetic spectrum exceeds the magnetised kinetic spectrum. While the kinetic energy in the unmagnetised case is sourced by truncation error which converges to zero, the initial varicose and kink instabilities which drive kinetic energy growth in the magnetised case are triggered by this same truncation error.  The numerical error between the kinetic and magnetic sectors is non-linearly coupled and so the magnitude of the kinetic energy in the unmagnetised case cannot be used as a simple proxy for the error in the magnetised kinetic spectrum.  

In the magnetic spectrum we see a very similar behaviour between Fig. \ref{fig:powerspec_f} and Fig. \ref{fig:powerspec_g}. When comparing the spectra normalised by the total magnetic field energy, we see that the magnetic spectra remains almost identical between ${\sim}$ 50 ms and the end of the simulation. If we neglect the overall dissipation of the magnetic energy, the spectral decomposition of the magnetic field appears to have reached a quasi-steady state, corresponding to a $k^{-5/3}$ spectrum. As in the kinetic spectrum case, this feature is clearer at the two higher resolutions compared to the lower resolutions.

In all plots discussed here we have rescaled the power spectrum by a factor of $k^{5/3}$, to compare with the K41 spectrum. We note that a rescaling by a factor of $k^{3/2}$ as argued by IK is also consistent with our findings, and, at larger wavelengths may even prove a marginally better fit.

\subsection{Helicity Evolution}

The helicity of the magnetic field measures the topological properties  of the field, and in ideal MHD is a conserved quantity \citep{Woltjer:1958}. Measurements of the deviation of the helicity from its initial value are therefore a clear indication of non-ideal effects, such as magnetic reconnection, or simply the loss of structure from the numerical simulation due to numerical dissipation. As the turbulence discussed above drives energy to smaller length scales, we can expect to see a growth of helicity at these times as the ideal MHD approximation is violated. We define (net) magnetic helicity as 
\bea
H = \int_V A\cdot B dV, \label{eq:helicity}
\eea 
where we will restrict our integral to the finest refinement level in our simulations, which completely covers the star. We can expect a small violation of conservation of helicity from the effect of choosing this finite domain, however, the magnetic field remains largely inside the star throughout the simulation. At the end of the timescales considered, the average field strength outside the finest refinement level is order $1\%$ of the field strength within the finest level, and is lower at earlier times. The helicity contribution in the exterior is therefore considerably smaller than that in the interior which dominates the discussion below.  In \AK the magnetic field $B$ is the evolved variable, and we do not evolve $A$. For the calculation of the helicity we reconstruct $A$ from $B$ using the publicly available cell-by-cell code of \cite{Silberman:2018ioy}. $A$ is not a gauge invariant quantity, and in the reconstruction of $A$ from $B$ the Coulomb gauge is enforced, following the approach of \cite{Silberman:2018ioy}. We note however that the helicity itself is gauge invariant, so long as the component of the magnetic field normal to the boundary of the integration volume in Eq. \ref{eq:helicity} vanishes. Again, this boundary lies in a region where the magnetic field strength is relatively low compared with the field inside the volume.

 \begin{figure}
 	\centering 
 	\includegraphics[width=0.49\textwidth]{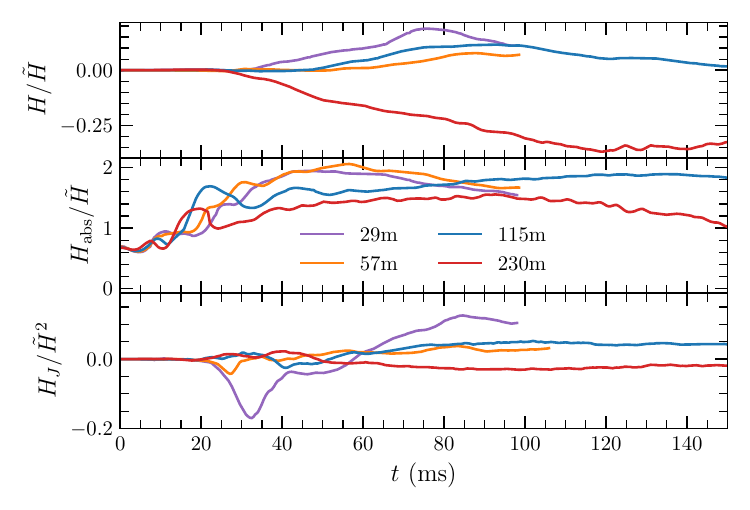}
	 \caption{Upper (Middle) panel: Evolution of net (absolute) helicity normalised by reference helicity. Lower panel: Evolution of current helicity.}
 	\label{fig:tothelicity}
 \end{figure}

 \begin{figure}
   \includegraphics[width=0.49\textwidth]{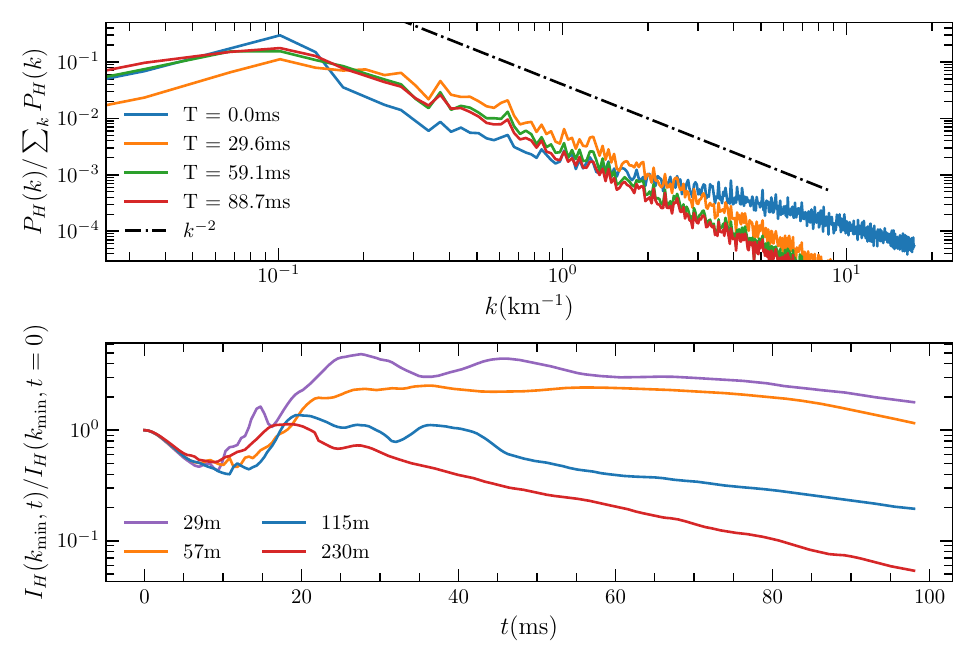}
   \caption{(Upper panel) Evolution of helicity spectrum normalised by total helicity. (Lower panel) Evolution of the Hosking invariant $I_H$ normalised by its initial value.}
 		\label{fig:helspec}
 \end{figure}

 In Fig. \ref{fig:tothelicity} (upper panel) we demonstrate the evolution of the net helicity. Following \cite{Ciolfi:2012en} we normalise the helicity by a reference helicity $\tilde{H}(t) = r_{NS}E_B(t)/2$, corresponding to a star with poloidal and toroidal fields in equipartition. At early times we see minimal growth in the helicity, except in the lowest resolution run. This feature, also visible in the simulations of \cite{Ciolfi:2012en} reduces considerably as a result of resolution, and at late times dissipates, as discussed below. At the higher three resolutions however, we see that the growth in helicity begins between $35-50$ ms into the simulation, coincident with the time of formation of the turbulent spectrum for that resolution, as discussed in Sec. \ref{sec:spectra}.
  At this moment, energy is cascading to smaller scales, before dissipating from the grid, leading to a growth in helicity.

 For the longer simulations, at late times we see that the net helicity appears to converge to zero. Since this configuration is nonrotating, with no initial toroidal field, there should be no preferred global orientation for the helicity to develop with. Hence we might expect that, integrated over the whole volume, the helicity should average to zero. It has further been noted by \cite{Braithwaite:2005md} that it may be possible to evolve to a non-trivial equilibrium configuration, the helicity of which exactly cancels to zero when integrated over the star. 
In \cite{Braithwaite:2005xi} a twisted-torus configuration was found as the equilibrium end state of the evolution of random initial magnetic field data, however for this end state to be reached the initial data must have finite helicity. The fact that our configurations do not appear to reach this end state (c.f. Fig. \ref{fig:stream_c}) is explained by a net helicity that appears to vanish at late times.
 
 To further investigate the helicity development, we demonstrate also the absolute helicity, 
\bea
H_{\mathrm{abs}} = \int_V |A\cdot B| dV,
\eea
	in Fig. \ref{fig:tothelicity} (middle panel). Note that we do not expect to see conservation of this absolute helicity, but we include it to demonstrate the effect of orientation cancellation in the net helicity described above. The largest absolute helicity is generated for the highest resolutions once the turbulent cascade has developed. These simulations resolve the cascade best, and so capture the most helicity generation at the smallest scales. At late times we do not see the absolute helicity decay to zero, showing that the dissipation of helicity is due to the global cancellation of helicity orientation. At higher resolutions the absolute helicity appears to converge to larger non-zero values, as the cascade forms. 
	This quantity provides a heuristic argument for the orientation of the helicity in
the star. Below we explore further the cancellation of the helicity orientation with a more
precise quantity, the Hosking invariant.

	We also consider the evolution of both the kinetic and current helicity, respectively $H_J = \mathbf{B} \cdot \left(\nabla \times \mathbf{B}\right)$, 
	$H_K = \mathbf{u} \cdot \left(\nabla \times \mathbf{u}\right)$. We find that the total kinetic helicity (not pictured) is dominated by numerical noise, 
	due to numerical effects arising at the surface of the star, but the current helicity (Fig. \ref{fig:tothelicity} lower panel) follows a similar structure to that of the magnetic helicity. 
	Best captured at highest resolution, we see a sharp increase in $H_J$ shortly after the saturation of the initial instabilities, peaking at 
	$\sim{}30$ms as the field becomes more and more twisted, with this feature continuing to grow in magnitude (though reversing its overall sign) 
	towards the end of the simulation, as the turbulent cascade develops. In stellar astrophysics it is observed that the helicity in the northern 
	and southern hemispheres of stars have opposite orientations. Our configuration is non rotating, and thus we may not expect such a distinction. 
	We split these helicities into their contributions in the northern and southern hemispheres of the star (not pictured) and find that, for all quantities 
	$H, H_K, H_J$ considered, both the northern and southern hemisphere quantities have the same sign, except for in our lowest resolution simulation, 
	where the signs differ. Further, we see that the northern and southern contributions are identical up until $\sim{}$20ms, at which point they begin to non 
	trivially differ, though this divergence is delayed as a function of resolution.

	In the presence of MHD turbulence it has been found that magnetic helicity will undergo an inverse cascade \citep{Frisch:1975} , as helicity is created at small scales and is then transferred to large scales. In Fig. \ref{fig:helspec} we demonstrate the spectrum of the magnetic helicity (defined in App. \ref{app:specdef}) at different times, normalised by the total helicity. In the initial data the helicity is dominated by the contribution from large scales, due to the lack of turbulence and overall large scale field. After $\sim 30$ ms more helicity has generated in an intermediate range, and the relative contribution of the large scale helicity has dropped, as the inverse cascade has not developed yet. By $\sim 60$ ms however, corresponding to the onset of the turbulent spectrum discussed in Sec. \ref{sec:spectra}, we can see that the helicity is transferring back to smaller values of $k$, and that this trend is continuing through to  $\sim 90$ ms, demonstrating an inverse cascade to large length scales. The specific nature of the power law scaling that the inverse cascade obeys is an open question, with various numerical studies suggesting exponents between $k^{-10/3}$ \citep{Mininni:2009} and $k^{-2}$ \citep{Pouquet:1976}. As demonstrated by the superposed dashed line in Fig. \ref{fig:helspec}, we find that $k^{-2}$ provides a good fit to the helicity scaling, while the $k^{-10/3}$ scaling provides a poor match.

In the case of turbulent flow with an initially zero helicity, \cite{Hosking:2020wom} have proposed a new integral invariant which should remain constant 
	while the magnetic field decays, in analogy with the Saffman invariant, which, in line with subsequent works, we refer to as the Hosking invariant, 
	$I_H(x) = \int d^3 r \left \langle H(x) H(x+r) \right \rangle$. Here angled brackets denote an ensemble average. The invariance of this quantity 
	is equivalent to the invariance of the squared helicity in the limit of an infinite volume. 

We can interpret the conservation of this quantity by considering its relationship to the magnetic Loitsyansky invariant $I_{L_M}(x) = - \int d^3 r r^2 \left \langle B(x) B(x+r) \right \rangle$. Since $I_{L_M}$ satisfies the relation $I_{L_M} \sim I_H / B^2$, the conservation of $I_H$ while the magnetic field strength decays implies an inverse cascade
	of magnetic field energy.

	We estimate the value of the Hosking invariant in our simulations, to verify its invariance, and provide evidence for such 
	an inverse cascade. \cite{Hosking:2020wom} show that, for small wavenumber $k$, the Hosking invariant is given by the power spectrum of the helicity, 
	$I_H = \lim_{k \rightarrow 0}P_H(k) $. In the lower panel of Fig. \ref{fig:helspec} we show the time evolution of $P_H(k_\mathrm{min})$, where 
	$k_\mathrm{min} = 1/(2R_{\mathrm{NS}})$ is the wavenumber associated to the diameter of the neutron star. This is the smallest value of $k$ 
	physically resolvable in our domain. Here we see that $I_H$  is well preserved for our higher resolution simulations. All configurations show 
	dynamics associated to the development of the initial instabilities up to $\sim{}20$ ms, after which our lower two resolutions demonstrate 
	a steady decay due to insufficient resolution, while the higher two resolution simulations preserve the invariant at approximately its initial value, 
	within a factor 2. This preservation further supports the idea of an inverse cascade present in our simulations.

\section{Conclusion}
\label{sec:con}

In this paper we have performed high resolution, long term simulations of magnetic field instabilities in the interior of NSs with the performance portable GRMHD code \AK. We successfully simulate with resolution down to 29m for 100ms and at lower resolutions simulate over 1.2 s of evolution. These are the highest resolution and longest lasting simulations to date for this problem, which have allowed us to, for the first time, capture the formation of a turbulent cascade in a NS interior, driven by magnetic field instabilities.

We observe the well documented initial linear instability phase \citep{Tayler:1957a, Tayler:1973a, Markey:1973a, Markey:1974a,
	Wright:1973a}, and
the subsequent non linear rearrangement of the magnetic field. 
We find that the toroidal field developed in the star
interior reaches $\sim 14\%$ of the total magnetic field energy in our
highest resolution simulation, and that on time scales of a second
saturates between $34-42\%$. The magnetic field dynamics triggers
oscillations in the star which we analyse on both large and small scales.

On the large scale we see the excitation of a long lived axisymmetric mode with a clear frequency of $145$ Hz.
This density oscillation is compatible with the presence of Alfven modes in the star, previously identified in \cite{Sotani:2007pp} but not yet observed in 3D GR simulations. These modes have been proposed as an explanation for QPOs seen in magnetars \citep{Israel:2005av}, though a full analysis will require the modeling of crustal elasticity.
On the small scale, the magnetic field triggers a clear turbulent cascade with scaling consistent with a Kolmogorov $k^{-5/3}$ power law.
This behaviour is absent for an unmagnetised star, but requires our highest resolution of 29m to be clearly captured with MHD.
In the magnetic field itself, we find that after 50ms of evolution the magnetic field spectrum reaches a quasi steady state relative to the overall magnetic field strength and consistent with this turbulent cascade. 

Associated to this turbulent cascade, we find that the onset of turbulence coincides with the growth of a net helicity in our system, representing the non ideal effects of dissipation at the smallest length scales. We also find that the absolute helicity of the magnetic field grows during our simulations, with the end state converging to a non-zero value at higher resolutions, suggesting that more of the turbulent cascade is captured at these resolutions. In contrast the net helicity appears to converge to zero at late times, suggesting that, in the absence of initial helicity or a preferred direction imposed by rotation, the helicity orientation cancels out and a twisted-torus is not formed.
The helicity itself undergoes an inverse cascade, with helicity produced during the turbulent phase on small length scales transported to larger scales throughout the simulation, with the inverse cascade following a $k^{-2}$ power law scaling.

In this work we have only considered a static star in the Cowling approximation. A more physically complete model of astrophysically relevant magnetised NSs must also include the effects of rotation and the full solution of the Einstein Equations. Work is ongoing to incorporate these effects into our simulations. This will allow us to determine the effects of rotation on the generation of toroidal fields and helicity, as well as its impact on Alfven modes in the star. Solving the Einstein equations for such configurations will also be key in determining the associated gravitational wave signals generated. In all cases, longer simulations and higher resolutions will be invaluable in assessing the magnitude and impact of turbulent flow on the overall star evolution, and the existence of any quasi-equilibrium configuration for the magnetic field in the star.

	\section*{Acknowledgements}
{   SB, RJ, BH acknowledge support for the MERLIN project. MERLIN is
   funded by the Deutsche Forschungsgemeinschaft (DFG) and the Narodowe
   Centrum Nauki (NCN) OPUS-LAP grant number 2022/47/I/ST9/01494, under the EU weave initiative. 
   SB acknowledges support by the DFG project ``Magnetfelddynamik in Neutronensternen Sternen'' MERLIN (MERLIN; BE 6301/6-1 Projektnummer: 524726453).  
   SB acknowledges support by the EU Horizon under ERC Consolidator Grant, no. InspiReM-101043372.
   JF was supported by the Department of Energy, Office of Science under award DE-SC0021177.
   This research used resources of the Argonne Leadership Computing Facility, which is a DOE Office of Science User Facility supported under Contract DE-AC02-06CH11357. An award of computer time was provided by the INCITE program.
   We acknowledge the EuroHPC Joint Undertaking for awarding this project access to the EuroHPC supercomputers, LEONARDO, hosted by CINECA (Italy) and the LEONARDO consortium; LUMI, hosted by CSC (Finland) and the LUMI consortium; and KAROLINA hosted by IT4Innovations National Supercomputing Center (Czech Republic) through an EuroHPC Benchmark Access call (EHPC-BEN-2024B10-018).
  Computations were also performed on the ARA cluster at Friedrich Schiller University Jena and on the {\tt Tullio} INFN cluster at INFN Turin. The ARA cluster is funded in part by DFG grants INST 275/334-1 FUGG and INST 275/363-1 FUGG, and ERC Starting Grant, grant agreement no. BinGraSp-714626. We thank Francesco Zappa for early contributions to this project and thank Axel Brandenburg, Aurora Capobianco, David Radice and James Stone for helpful comments and discussions.}

\section*{Data Availability}
	The data used to generate the figures for this paper is available from the authors upon reasonable request.

\bibliographystyle{mnras}

\appendix

\section{Boundary conditions}
\label{sec:BC}

In order to assess the robustness of our simulations we perform a series of runs, varying the size of the outer boundary, to investigate the impact of boundary effects on our simulations. With the resolution on the star fixed to $\delta x  = 230$m, we vary the distance of the outer boundary from the origin, $L$, from (29km, 59km, 118km, 235km).

\begin{figure}
\centering 
\includegraphics[width=0.49\textwidth]{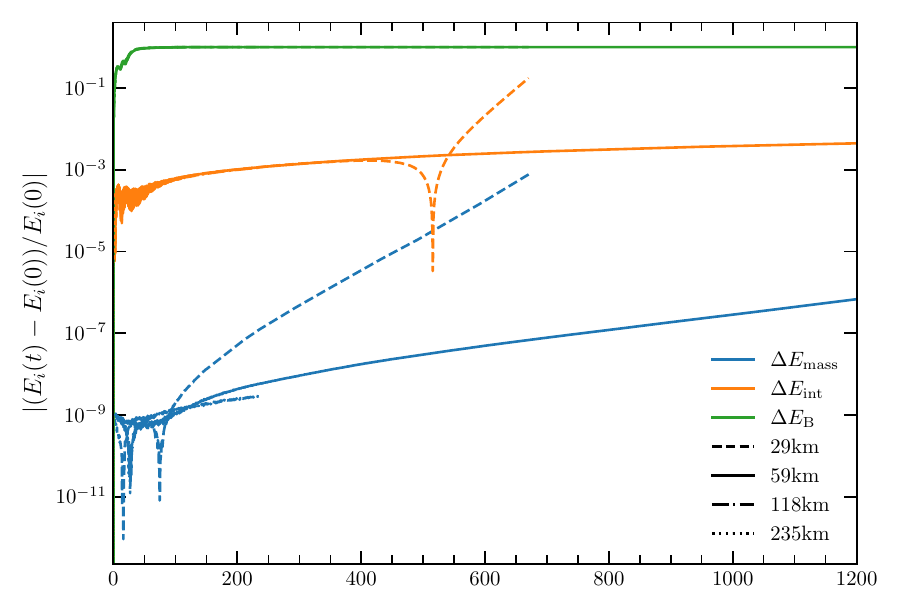}
\caption{ Relative change in a given energy sector $\Delta E_i$, compared to its initial value $E_i(t=0)$, shown for rest mass energy, internal energy and magnetic energy. Linestyles demonstrate different distances of outer boundary $L$.}
\label{fig:boundary}
\end{figure}

In Figure \ref{fig:boundary} we demonstrate the evolution of the relative energy change $\Delta E_i = \left| (E_i(t) - E_i(0))/E_i(0)\right|$ for $E_{\mathrm{int}}, E_{\mathrm{B}}, E_{\mathrm{mass}}$. Clearly $L=29$km gives the least accurate result, with the error in the conserved rest-mass growing to ${\sim} 0.1\%$ by $\sim700$ms, and performing considerably worse than all other boundary conditions after ${\sim}100$ms. After ${\sim}500$ms the internal energy also becomes unstable, reaching an error of ${\sim}10\%$ by $\sim700$ms. Unlike the other runs, which have a small dissipation of the internal energy on the level of ${\sim}0.1\%$, the $L=29$km boundary has a growing internal energy after ${\sim}500$ms. The $L=59$km boundary however remains stable for the full 1.2s length of our simulation, both in the internal energy and restmass. While the larger boundary conditions shown do have a better preservation of the rest mass, by ${\sim}200$ms the $L=59$km run develops a relative error of only ${\sim}10^{-6}$ which we consider sufficiently robust for our results above. We note that in \cite{Sur:2021awe} an outer boundary $L=29$km was used, potentially explaining the poor conservation of internal energy seen in that work. We also note that the overall dissipation of magnetic field energy is a feature seen in all simulations, and is thus not a result of the choice of boundary condition.  

\section{Energy decomposition}
\label{app:energies}
We consider a perfect magnetised fluid in the ideal MHD approximation, with stress energy tensor $T^{\mu\nu} = (\rho h+b^2) u^\mu u^\nu + (p + \frac{b^2}{2})g^{\mu\nu} - b^\mu b^\nu$, where $\rho$ is the fluid density, $h$ enthalpy, $u^\mu$ 4-velocity and $b^\mu = u_\nu{}^*F^{\mu\nu}$ the magnetic field as seen by the comoving fluid observer.

In the 3+1 decomposition of the Euler equations the energy density seen by an Eulerian observer moving along the normal vector to the foliation $n^\mu$ is given by $E = n_\mu n_\nu T^{\mu\nu}$. In the so-called Valencia formulation of \cite{Banyuls:1997zz}, this is replaced as an evolution variable by $\tau = E- D$, where $D = \rho W$ is the mass-density seen by the Eulerian observer, and $W$ is the Lorentz factor. We characterise the different contributions to the energy budget of our simulations though the decomposition of this $\tau$ as follows. Expressed in primitive variables, 

\begin{eqnarray}
\tau 
&=& \rho W (W-1) + W^2 \rho \epsilon + p(W^2 - 1) + b^2\left(W^2 - \frac{1}{2}\right) - (\alpha b^0)^2, \nonumber \\	
\end{eqnarray}

where $\alpha$ is the lapse function.

We identify,

\bea
E_{\mathrm{KE}} &=& \rho W (W-1),\\
E_{\mathrm{int}} &=&  W^2 \rho \epsilon + p(W^2 - 1),\\
E_{\mathrm{B}} &=&  b^2\left(W^2 - \frac{1}{2}\right) - (\alpha b^0)^2,\\
E_{\mathrm{mass}} &=& D,\\
\tau &=& E_{\mathrm{KE}} + E_{\mathrm{int}} + E_{\mathrm{B}}.
\eea

In the Valencia formulation neither $E$ nor $\tau$ are conserved variables, they satisfy balance law equations with source terms arising due to the presence of a non-trivial metric. 
While these sources contain gauge dependent terms, they also capture the coupling of the momentum density to the gravitational potential. 
By artificially holding the metric constant through the Cowling approximation we provide an unphysical value for the gravitational field in this source term, 
though due to the small size of the magnetic field energy in comparison with the rest mass energy of the NS, we expect this effect to be small.

\section{Alfven Time} 
\label{app:alf}

We define the Alfven time associated to the magnetised star as

\bea
\tau_A &=& \frac{2R_{\mathrm{NS}}}{v_A},\\
v_A &=& \frac{\langle B \rangle}{\langle\sqrt{4\pi\rho}\rangle},
\eea

where $\langle f \rangle$ denotes the volume average of $f$ over the star interior.
The Alfven time calculated from the initial data of our configuration is ${\sim}13$ms. In terms of this initial Alfven time, our simulations range between 7.7 and 92 Alfven times in length. Since the magnetic field strength rapidly decays over the course of the simulation however, the Alfven time grows over the course of the simulation. We define the cumulative Alfven time elapsed during the simulation as 

\bea
T_A = \int^t_0 \frac{dt'}{\tau_A(t')}.
\eea

For the configurations considered, we find that, after 100ms have elapsed $T_A \approx 2.5$, while, by the end of the full 1.2s simulation at 230m resolution $T_A\approx 3.5$, considerably shorter than estimated using the initial Alfven time. We note that other studies, for instance \cite{Tsokaros:2021pkh}, define the Alfven time in terms of the central magnetic field and maximum density using $\tau_A = R_{\mathrm{NS}} \sqrt{4\pi\rho_{\mathrm{max}}}/B_{\mathrm{c}}$. To facilitate easy comparison between different studies in the literature using different conventions, we recalculate our Alfven time also with this definition. With this convention our initial Alfven time is ${\sim} 3.3$ms. Therefore, under this definition our highest resolution simulation lasts for 30.4 Alfven times, while our longest simulation lasts for 365 Alfven times.

\section{Power Spectra}
\label{app:specdef}
We here define the diagnostic quantities used to  analyse the power spectra in Sec. \ref{sec:spectra}. We follow standard convention for analysis of turbulent phenomena, see e.g. \cite{Pope:2000}. We define the Fourier transform of a function $f$, denoted $\tilde{f}$, by

\begin{eqnarray}
	\tilde{f}(\mathbf{k}) = \int f(\mathbf{x}) e^{-2\pi i \mathbf{x}\cdot\mathbf{k}} d^3x.
\end{eqnarray}

In practice, we calculate Fourier transforms over the finest refinement level in the simulation, which fully covers the star, and window $f$ to zero outside the star surface to ensure a periodic signal for the numerical discrete Fourier transform.

We define the power spectra, associated to the specific kinetic energy, $P_{v}(k)$, specific magnetic field energy $P_{b}(k)$ and helicity $P_{H}(k)$. 
\begin{eqnarray}
	P_{v}(k_r) &=& \int_{\Omega_k} P_{v}(\mathbf{k}) d\Omega_k =\int_{\Omega_k} \left(\tilde{v^i}(\mathbf{k})\right)^{\dagger} \tilde{v_i}(\mathbf{k}) d\Omega_k \\
	P_{b}(k_r) &=& \int_{\Omega_k} P_{b}(\mathbf{k}) d\Omega_k \\ 
	&=& \int_{\Omega_k} \left(\tilde{\frac{B^i}{\sqrt{\rho}}}(\mathbf{k})\right)^\dagger \tilde{\frac{B_i}{\sqrt{\rho}}}(\mathbf{k}) d\Omega_k \nonumber\\
	P_H(k_r) &=& \int_{\Omega_k} P_{H}(\mathbf{k}) d\Omega_k\\
	&=& \int_{\Omega_k} \frac{1}{2}\left(\left(\tilde{A}^{i}(\mathbf{k})\right)^\dagger \tilde{B_i}(\mathbf{k}) + \tilde{A}^{i}(\mathbf{k}) \left(\tilde{B_i}(\mathbf{k})\right)^{\dagger} \right) d\Omega_k \nonumber
	\label{eq:powerspec}
\end{eqnarray}
We note that the first two of these quantities  are not the power spectra associated to the energies defined in Appendix \ref{app:energies} and discussed in Sec.~\ref{sec:energy}, however these quantities provide a natural comparison with quantities commonly analysed in studies of Newtonian MHD turbulence, and quantities arising in the power spectra associated to the Newtonian Els\"asser variables, which naturally generalise to general relativity as shown by \cite{Chandran:2017zdg}. We choose to analyse the magnetic field densitised in this way such that we are analysing a specific energy as in the kinetic sector. This is equivalent to analysing the magnetic field in Alfvenic units. 
Here the domain of integration $\Omega_k$ is a sphere of radius $k$ in Fourier space, with the area element $d \Omega_k  = k^2 \sin(k_\theta)d k_\theta dk_\phi$, with $k_\theta,k_\phi$ the standard colatitudinal and azimuthal angles of the spherical polar coordinate system defined in Fourier space, $(k_r,k_\theta,k_\phi) $. The power spectra that are presented are therefore functions of the radial wavenumber $k_r$, with the angular dependence integrated over.

\bsp
\end{document}